\documentclass[aps,pra,preprint,groupedaddress]{revtex4-2}

\usepackage[dvips]{graphicx}
\usepackage{amsmath,amssymb}
\usepackage{mathrsfs}

\begin{document}


\title{Long time self-modulation of nonlinear electromagnetic wave 
in two-dimensional cavity}


\author{Kazunori Shibata}
\affiliation{Institute of Laser Engineering, Osaka University, 2-6 Yamada-Oka, Suita, Osaka 565-0871 Japan}


\date{\today}

\begin{abstract}
The vacuum is expected to exhibit electromagnetic nonlinearity. 
We demonstrate the properties of nonlinear electromagnetic wave 
in a two-dimensional rectangular cavity by calculating the 
nonlinear correction for two classical standing waves. 
We apply the linear approximation 
in a short timescale. A part of the nonlinear correction increases 
with time. In particular, a one-dimensional second harmonic 
grows if the cavity size satisfies a specific condition. 
We also analyze the nonlinear electromagnetic wave in 
a timescale longer than the applicable limit of the linear 
approximation. 
We formulate the self-modulation of the amplitude 
and phase, including the effect of 
static magnetic flux density. 
In the viewpoint of energy flow between the two modes of 
the standing wave, the behavior of nonlinear electromagnetic wave 
can be classified into three types. 
Namely, the energy flow keeps oscillating, 
eventually decreases to zero, or never occurs. 
\end{abstract}


\maketitle

\section{Introduction}

An electromagnetic field in the classical vacuum is 
well described by the linear Maxwell's equations. 
After the advent of quantum field theory in the 20th century, 
a correction to the classical electromagnetic field has emerged. 
For example, a correction by 
virtual pairs of electron and position in 
quantum electrodynamics is known as the 
Heisenberg-Euler model 
\cite{Heisenberg1936}\cite{PhysRev.82.664}. 
Another famous model is the Born-Infeld model 
\cite{doi:10.1098-rspa.1934.0059} 
which is derived by an analogy to the theory of relativity. 
These corrections yield nonlinear Maxwell's equations.

Such a nonlinear correction has been pointed out to 
affect various phenomena, such as 
the Wichmann-Kroll correction
to the Lamb shift \cite{PhysRev.101.843} and  
a correction to the energy levels of a hydrogen atom 
\cite{Denisov2006}\cite{PhysRevLett.96.030402}\cite{Mazharimousavi2012}\cite{Akmansoy2018}. 
However, an experimental verification has yet to succeed.

Various experiments and experimental proposals 
have been designed 
to verify the electromagnetic vacuum nonlinearity.  
A focusing of a strong laser beam is 
typically considered \cite{RRP-S145}. 
Such an attempt aims to generate a nonlinear effect 
by instantaneously achieving an extremely large intensity
\cite{RevModPhys.84.1177}\cite{king_heinzl_2016}
\cite{RevModPhys.78.309}. 
However, a strong laser is not the only approach. 
For example, nonlinear behaviors in a cavity system 
\cite{PhysRevA.70.013808}\cite{PhysRevLett.87.171801}
\cite{Vlasov2015}\cite{PhysRevD.97.096005}
\cite{PhysRevA.63.012107}
\cite{Shibata2020}\cite{Shibata2021EPJD}, 
a waveguide \cite{PhysRevLett.87.171801}\cite{Brodin_2002PS}
\cite{PhysRevLett.99.230401}, and  
a ring laser \cite{Denisov2001} 
have been calculated. 
In particular, several experiments using a cavity have been 
performed to detect vacuum birefringence 
\cite{PhysRevD.90.092003}\cite{DELLAVALLE20104194}\cite{Zavattini_2013}
\cite{PhysRevD.78.032006}\cite{DellaValle2016}
\cite{Cadene2014}\cite{Fan2017}.

A cavity is capable of confining an electromagnetic wave 
in a long time compared to the timescale of laser focusing. 
A characteristic behavior of nonlinear electromagnetic 
wave can appear in a long timescale 
by accumulating an instantaneously small nonlinear effect, 
as reported in a one-dimensional system 
\cite{PhysRevA.104.063513}. 
However, it is hard to predict the behavior of 
nonlinear electromagnetic waves in a 
two- or three-dimensional cavity 
because a physical phenomenon generally changes its behavior 
depending on the spatial dimension. 
Thusly, it is worth clarifying theoretically the property of 
nonlinear electromagnetic waves for future experiments.

In this study, we analyze a nonlinear electromagnetic wave 
in a two-dimensional rectangular cavity. 
First, we employ the linear approximation and clarify the 
condition that the nonlinear corrective term 
can increase with time. 
Then, we elucidate the leading term of 
nonlinear electromagnetic waves in a longer timescale 
than the applicable range of the linear approximation. 
As a characteristic behavior 
in the two-dimensional cavity, 
we report that a second harmonic can 
increase with time depending on the cavity size.

\section{Notation, system, and classical term}
The electromagnetic fields are normalized by 
the electric constant $\varepsilon_0$ and 
magnetic constant $\mu_0$ as follows. 
The electric field $\boldsymbol{E}$ and 
vacuum polarization $\boldsymbol{P}$ are 
multiplied by $\varepsilon_0^{1/2}$ and 
$\varepsilon_0^{-1/2}$, respectively. Similarly, 
the magnetic flux density $\boldsymbol{B}$ and 
vacuum magnetization $\boldsymbol{M}$ are 
multiplied by $\mu_0^{-1/2}$ and 
$\mu_0^{1/2}$, respectively. 
We suppose $\boldsymbol{E}$ and $\boldsymbol{B}$ 
to be of class $C^1$.

We consider the simplest nonlinear Lagrangian density $L$ 
in the Pleba{\'{n}}ski class \cite{Plebanski1970} as  
\begin{equation}
L=\frac{1}{2}F+C_{2,0}F^2+C_{0,2}G^2. 
\end{equation}
This form is frequently used because it can be an effective 
Lagrangian if the electromagnetic fields are not extremely strong. 
$C_{2,0}$ and $C_{0,2}$ are the nonlinear parameters, 
{\it e.g.}, their values in the Heisenberg-Euler model are 
$C_{2,0}=1.665\times 10^{-30}$(m$^3$/J) and $C_{0,2}=7C_{2,0}$, 
respectively\cite{PhysRev.82.664}\cite{PhysRevD.93.093020}
\cite{PhysRevD.98.056001}. 
The polarization $\boldsymbol{P}$ and 
magnetization $\boldsymbol{M}$ of vacuum are 
defined as 
\begin{equation}
\begin{split}
&\boldsymbol{P}=
4C_{2,0}F\boldsymbol{E}+2C_{0,2}G\boldsymbol{B},\\
&\boldsymbol{M}=
-4C_{2,0}F\boldsymbol{B}+2C_{0,2}G\boldsymbol{E},\\
\end{split}
\end{equation}
respectively. 
The charge $\rho$ and current $\boldsymbol{j}$ of vacuum 
are composed of electromagnetic field itself as 
\begin{equation}\label{hagen_rho_j}
\begin{split}
&\rho=-\nabla\cdot\boldsymbol{P},\\
&\boldsymbol{j}=c^{-1}\partial_t\boldsymbol{P}
+\nabla\times\boldsymbol{M},\\
\end{split}
\end{equation}
where $c$ is the speed of light and 
$\partial_t$ denotes the partial differentiation 
with respect to time $t$. 
The nonlinear Maxwell's equations are given by 
\begin{equation}\label{eq_nme}
\begin{split}
&\nabla\cdot\boldsymbol{B}=0,\\
&\nabla\times\boldsymbol{E}+
c^{-1}\partial_t \boldsymbol{B}=
\boldsymbol{0},\\
&\nabla\cdot\boldsymbol{E}=\rho,\\
&\nabla\times\boldsymbol{B}-
c^{-1}\partial_t\boldsymbol{E}=\boldsymbol{j}.\\
\end{split}
\end{equation}
The nonlinearity appears in the form of $\rho$ and $\boldsymbol{j}$.

The physical system we treat is a 
two-dimensional cavity whose domain is set to 
$0\le x \le \ell_1, 0 \le y \le \ell_2$ and 
the boundary is supposed to be a perfect conductor mirror. 
On the surface of the mirror, a static magnetic flux density 
$\boldsymbol{B}_s=(B_{sx},B_{sy},B_{sz})$ can exist. 
The boundary conditions are given by 
\begin{equation}\label{bc_EB}
\begin{split}
&E_y(0,y,t)=0, \ \ E_y(\ell_1,y,t)=0, \\
&E_z(0,y,t)=0, \ \ E_z(\ell_1,y,t)=0, \\
&B_x(0,y,t)=B_{sx}, \ \ B_x(\ell_1,y,t)=B_{sx}, \\
&E_x(x,0,t)=0, \ \ E_x(x,\ell_2,t)=0, \\
&E_z(x,0,t)=0, \ \ E_z(x,\ell_2,t)=0, \\
&B_y(x,0,t)=B_{sy}, \ \ B_y(x,\ell_2,t)=B_{sy}. \\
\end{split}
\end{equation}

Here we describe a classical standing wave that can exist 
in the cavity. Let $k>0$ be the magnitude of the wave vector. 
The wave direction is 
expressed by an angle $\theta$. Using two 
natural numbers $n_1$ and $n_2$, $\theta$ satisfies 
$\cos \theta=n_1\pi/(k\ell_1)$ and 
$\sin \theta=n_2\pi/(k\ell_2)$ because of the boundary conditions. 
The frequency is given by $\omega=ck$. 
There are two modes of the standing wave with this wave vector. 
Let $A_1, A_2 \ge 0$ be the amplitudes of respective modes 
and $\Phi$ be the relative phase. 
At least one of $A_1$ or $A_2$ are supposed to be nonzero. 
We abbreviate to $X=(k\cos\theta)x, Y=(k\sin\theta)y,$ 
and $T=\omega t$. 
The classical electromagnetic fields 
$\boldsymbol{E}_c$ and $\boldsymbol{B}_c$ are 
generated at a certain negative time and 
their values at $t\ge 0$ are given by 
\begin{equation}\label{koten_EcBc}
\begin{split}
&\boldsymbol{E}_c=
\begin{pmatrix}
-A_1 \sin \theta \cos X \sin Y \sin T\\
A_1 \cos \theta \sin X \cos Y \sin T\\
A_2 \sin X \sin Y \sin (T+\Phi)\\
\end{pmatrix}, \\
&\boldsymbol{B}_c=
\begin{pmatrix}
A_2 \sin \theta \sin X \cos Y \cos(T+\Phi)\\
-A_2 \cos \theta \cos X \sin Y \cos(T+\Phi)\\
A_1 \cos X \cos Y \cos T\\
\end{pmatrix}
+\boldsymbol{B}_s,  \\
\end{split}
\end{equation}
where we set $\boldsymbol{B}_s$ to be constant. 
These fields satisfy the classical linear Maxwell's equations. 
However, they do not necessarily satisfy the 
nonlinear Maxwell's equations. The difference between the 
total electromagnetic field and the classical term is 
referred to as the corrective term and expressed by 
a subscript $n$. 
Thus, we can express as 
$\boldsymbol{E}=\boldsymbol{E}_c+\boldsymbol{E}_n$ 
and $\boldsymbol{B}=\boldsymbol{B}_c+\boldsymbol{B}_n$. 
Our concern is to calculate the corrective term, 
in particular, its magnitude. 

We first apply a linear approximation by assuming that 
the corrective term is much smaller than the classical term. 
The corrective term within the range of the 
linear approximation is especially called 
``minimum corrective term'' and we 
attach a superscript $(0)$ in addition to a subscript $n$. 
The minimum corrective term 
is the first-order correction of the regular perturbation. 
We expect that the minimum corrective term is 
a good approximation of the exact corrective term 
in a short timescale, 
{\it i.e.}, $\boldsymbol{E}_n^{(0)} \approx \boldsymbol{E}_n$. 
In the following calculations, 
at least one of $B_{sx}$ and $B_{sy}$ is zero 
to avoid the situation of no solution \cite{Shibata_2022}.

\section{Linear approximation}

In the linear approximation, 
the charge and current in Eq. (\ref{hagen_rho_j}) are 
composed only of the classical term. 
We express them as $\rho_c$ and 
$\boldsymbol{j}_c$, respectively. 
The minimum corrective term is 
generated by these wave sources and satisfies 
the following equations: 
\begin{equation}\label{eq_mct}
\begin{split}
&\nabla\cdot\boldsymbol{B}_n^{(0)}=0,\\
&\nabla\times\boldsymbol{E}_n^{(0)}+
c^{-1}\partial_t \boldsymbol{B}_n^{(0)}=
\boldsymbol{0},\\
&\nabla\cdot\boldsymbol{E}_n^{(0)}=\rho_c,\\
&\nabla\times\boldsymbol{B}_n^{(0)}-
c^{-1}\partial_t\boldsymbol{E}_n^{(0)}=\boldsymbol{j}_c.\\
\end{split}
\end{equation}
The boundary conditions for the minimum corrective term 
are given according to the conditions in Eq. (\ref{bc_EB}).

The minimum corrective term is a 
sum of the homogeneous and special solutions 
of Eq. (\ref{eq_mct}). 
In the framework of classical 
electromagnetism, the homogeneous solution 
is proven not to increase with time. Furthermore, the 
initial distribution of minimum corrective term affects only 
the homogeneous solution. 
Therefore, it is sufficient to elucidate the behavior of 
the special solution to discuss the magnitude of the 
minimum corrective term. 

\subsection{Resonant increase}
A part of the minimum corrective term resonantly 
increases with time. We express such a part by 
$\boldsymbol{E}_{\text{reso}}$ and 
$\boldsymbol{B}_{\text{reso}}$. By defining constants 
\begin{equation}
\begin{split}
&\Gamma=\frac{1}{8}C_{2,0}
(4-\sin^22\theta),\\
&\Gamma_1=4C_{2,0}B_{sz}^2+C_{0,2}
(B_{sx}^2\sin^2\theta+B_{sy}^2\cos^2\theta),\\
&\Gamma_2=4C_{2,0}(B_{sx}^2\sin^2\theta+B_{sy}^2\cos^2\theta)
+C_{0,2}B_{sz}^2,\\
&\tilde{\Gamma}=-\frac{1}{8}\left(3C_{2,0}\sin^22\theta
-C_{0,2}\right),\\
\end{split}
\end{equation}
and using 
\begin{equation}\label{fgpq1}
\begin{split}
&f_1=(\Gamma-\tilde{\Gamma})A_1A_2^2 \sin \Phi \cos \Phi,\\
&g_1=-\Gamma_1A_1-3\Gamma A_1^3
-\Gamma A_1A_2^2 \cos^2 \Phi
-\tilde{\Gamma}A_1A_2^2 \sin^2 \Phi,\\
&p_1=\left(\Gamma_2+3\Gamma A_2^2
+\tilde{\Gamma}A_1^2\right)A_2 \sin \Phi,\\
&q_1=-\left(\Gamma_2+3\Gamma A_2^2
+\Gamma A_1^2\right)A_2 \cos \Phi,\\
\end{split}
\end{equation}
we obtain 
\begin{equation}\label{reso_mct_EB}
\begin{split}
&\boldsymbol{E}_{\text{reso}}=T
\begin{pmatrix}
- \sin \theta(f_1 \sin T+g_1 \cos T) \cos X \sin Y\\
 \cos \theta(f_1 \sin T+g_1 \cos T) \sin X \cos Y\\
(p_1 \sin T+q_1 \cos T) \sin X \sin Y\\
\end{pmatrix},\\
&\boldsymbol{B}_{\text{reso}}=T
\begin{pmatrix}
 \sin \theta(-q_1 \sin T+p_1 \cos T) \sin X \cos Y\\
- \cos \theta(-q_1 \sin T+p_1 \cos T) \cos X \sin Y\\
(-g_1 \sin T+f_1 \cos T) \cos X \cos Y\\
\end{pmatrix}.\\
\end{split}
\end{equation}
The resonant terms are partially calculated 
in Ref. \cite{Shibata2021EPJD}. 
The behavior that is proportional to time $T$ is the same 
as the resonance discussed in a one-dimensional system 
\cite{Shibata2020}\cite{Shibata2021EPJD}. 
One can see that there is an upper limit in the 
applicable time of the linear approximation because 
the resonant terms in Eq. (\ref{reso_mct_EB}) 
must be much smaller than 
the classical term.

\subsection{Another increase}

In the two-dimensional system, there can be an 
increasing solution which possesses a different property. 
The essential origin is a spatially uniform current. 
In the present system, a part of $\boldsymbol{j}_c$ is given by 
\begin{equation}\label{j_Ckl}
\boldsymbol{j}_{\text{uni}}=-\frac{k}{4}(4C_{2,0}-C_{0,2})A_1A_2
\sin 2\theta \cos (2T+\Phi)
\begin{pmatrix}
B_{sy}\\
B_{sx}\\
0\\
\end{pmatrix}, 
\end{equation}
where the corresponding charge is zero. 
All other terms in $\boldsymbol{j}_c$ depend on $x$ or $y$. 
Such a uniform current is characteristic to 
electromagnetic nonlinear interaction, in other words, 
it cannot be realized by a matter. 
The corresponding minimum corrective term must satisfy 
the boundary conditions and be bounded at $T=0$. 
Such a solution is obtained as 
\begin{equation}\label{EB_uni_cur_2D}
\begin{split}
&E_x=B_{sy}\left[-\frac{1}{2}(1-\cos 2ky)\sin (2T+\Phi)
+K_1(T-ky)-K_1(T+ky)\right],\\
&E_y=B_{sx}\left[-\frac{1}{2}(1-\cos 2kx)\sin(2T+\Phi)
+K_2(T-kx)-K_2(T+kx)\right],\\
&
\begin{split}
B_z=&B_{sx}\left[-\frac{1}{2}\sin2kx \cos(2T+\Phi)
+K_2(T-kx)+K_2(T+kx)\right]\\
&-B_{sy}\left[-\frac{1}{2}\sin 2ky \cos (2T+\Phi)
+K_1(T-ky)+K_1(T+ky)\right],\\
\end{split}\\
\end{split}
\end{equation}
$E_z=0,B_x=0$, and $B_y=0$, 
where the common coefficient 
$-(1/4)(4C_{2,0}-C_{0,2})A_1A_2\sin2\theta$ is omitted. 
The function $K_1(\bar{T})$ is 
defined at $\bar{T}\ge-k\ell_2$ and given as 
\begin{equation}\label{K1_tau_zenbu}
K_1(\bar{T})=
\begin{cases}
-\frac{1}{4}\left[\sin \left(2\bar{T}+\Phi\right)- \sin \Phi\right]
+\frac{1}{2}\left[
\tilde{B}_1\left(-k^{-1}\bar{T}\right)
+\tilde{E}_1\left(-k^{-1}\bar{T}\right)
\right]
& \left(-k\ell_2 \le \bar{T} \le 0\right)\\
\frac{1}{4}\left[\sin \left(2\bar{T}+\Phi\right)-\sin \Phi \right]
+\frac{1}{2}\left[
\tilde{B}_1\left(k^{-1}\bar{T}\right)
-\tilde{E}_1\left(k^{-1}\bar{T}\right)
\right]
& \left(0 \le \bar{T} < k\ell_2\right)\\
K_1\left(\bar{T}-2k\ell_2\right)
-\frac{1}{2}(1-\cos 2k\ell_2)
\sin \left(2\bar{T}-2k\ell_2+\Phi\right)
& \left(\bar{T} \ge k\ell_2\right),\\
\end{cases}
\end{equation}
where functions $\tilde{E}_1(y)$ and $\tilde{B}_1(y)$ 
express the initial distribution of the special solution 
and their domain is $y \in [0, \ell_2]$. 
The function $K_2(\bar{T})$ is defined at $\bar{T}\ge-k\ell_1$ 
and is obtained by replacing 
$\tilde{E}_1, \tilde{B}_1,$ and $\ell_2$ 
in the above equation 
by $\tilde{E}_2, \tilde{B}_2,$ and $\ell_1$, respectively. 
The domain of $\tilde{E}_2(x)$ and $\tilde{B}_2(x)$ 
is $x \in [0, \ell_1]$. 
These $\tilde{E}_1(y), \tilde{B}_1(y), \tilde{E}_2(x)$, 
and $\tilde{B}_2(x)$ are supposed to be of class $C^1$ and 
satisfy 
\begin{equation}\label{}
\begin{split}
&\tilde{E}_1(0)=\tilde{E}_1(\ell_2)=0,\\
&\tilde{B}_1'(0)=\tilde{B}_1'(\ell_2)=-k \cos \Phi,\\
&\tilde{E}_2(0)=\tilde{E}_2(\ell_1)=0,\\
&\tilde{B}_2'(0)=\tilde{B}_2'(\ell_1)=-k \cos \Phi.\\
\end{split}
\end{equation}
With these boundary values, we can confirm that 
$K_1, K_2,$ and the minimum corrective term 
to be of class $C^1$. 
Furthermore, if 
$\tilde{E}_1(y), \tilde{B}_1(y), \tilde{E}_2(x)$, 
and $\tilde{B}_2(x)$ are of class $C^2$ and satisfy 
$\tilde{E}_1''(0)=\tilde{E}_1''(\ell_2)=
\tilde{E}_2''(0)=\tilde{E}_2''(\ell_1)=-2k^2\sin \Phi$, 
then $K_1, K_2$, and the minimum corrective term are 
of class $C^2$.

\begin{figure}
\begin{center}
\begin{tabular}{c}
\includegraphics[width=8cm]{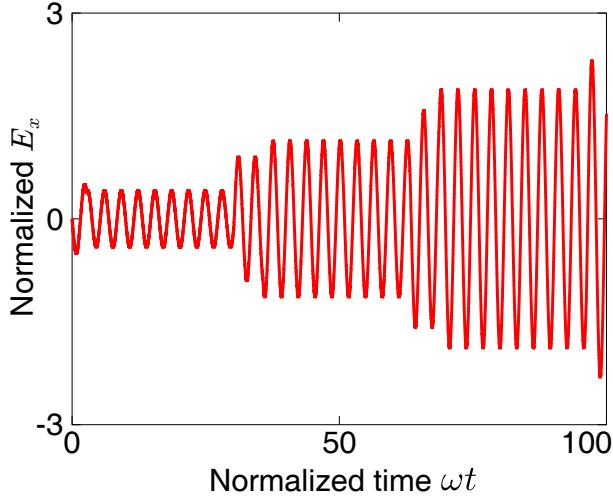}  \\
\end{tabular}
\end{center}
\caption{
Example of increasing second harmonic generated by 
the uniform current in the case of $\cos k\ell_2=0$. 
The electric field $E_x$ is normalized by 
$-(1/4)(4C_{2,0}-C_{0,2})A_1A_2B_{sy}
\sin2\theta$. 
The parameters are set to 
$k\ell_2=10.5\pi, ky=2.718$, and $\Phi=0$. 
The graph is continuous but not differentiable at 
$\omega t=\pm kx+nk\ell_2, n \in\mathbb{N}$ because 
we employ $\tilde{E}_1=0$ and $\tilde{B}_1=0$. 
\label{uni_j_coskl}}
\end{figure}

It will be worth noting that the minimum corrective term 
generated by the uniform current has a property that 
its time evolution varies with the cavity size. 
We fix $y$ and demonstrate the time-evolution of $E_x$ 
in the case of $B_{sy}\neq0$. 
Using a certain $t_0 \in [0,2c^{-1}\ell_2)$ and a 
natural number $n$, 
a general time $t\ge 0$ can be expressed as 
$t=t_0+2nc^{-1}\ell_2$. Thus, we obtain 
\begin{equation}\label{}
\begin{split}
E_x(y,t)=&E_x(y,t_0)-\frac{B_{sy}}{2}(1-\cos 2ky)
[\sin(2\omega t_0+4nk\ell_2+\Phi)-\sin(2\omega t_0+\Phi)]\\
&+B_{sy}(1-\cos 2k\ell_2)
\sum_{p=1}^n\cos[2\omega t_0+(4p-2)k\ell_2+\Phi]\sin 2ky,\\
\end{split}
\end{equation}
where the sum originates from the relationship in 
Eq. (\ref{K1_tau_zenbu}). 
The first and second terms in the right-hand side 
do not increase with $n$, {\it i.e.}, they are bounded 
with respect to time. 
However, the sum in the third term can increase with $n$. 
In fact, if $\cos k\ell_2=0$, then the 
third term becomes 
$2nB_{sy}\cos(2\omega t_0-2k\ell_2+\Phi)\sin 2ky$, 
which clearly increases with $n$. 
Figure \ref{uni_j_coskl} shows an example. 
If $\cos k\ell_2\neq0$, the third term is also bounded 
with respect to time.

Similar calculation shows that 
$E_y$ can increase with time 
in the case of $\cos k\ell_1=0$. 
The one-dimensional 
second harmonic wave generated by the 
uniform current can increase with time only if 
the cavity size and wave number satisfy the specific 
conditions that $\cos k\ell_1=0$ or $\cos k\ell_2=0$. 
Such a behavior is obviously different from the 
resonant increase.

\subsection{Applicable time of linear approximation}

The linear approximation 
can be a good approximation only when the minimum corrective 
term is much smaller than the classical term. 
All terms of minimum corrective term 
other than abovementioned do not increase with time. 
The amplitude of resonantly increasing term is 
about $(C_{2,0}+C_{0,2})(A_1+A_2+B_s)^3\omega t$. 
Similarly, that of the increasing term generated by 
the uniform current is about 
$(C_{2,0}+C_{0,2})(A_1+A_2+B_s)^3(c/\ell_1)t$, 
in the case of $\cos k\ell_1=0$. 
If we choose a typical size of the cavity as 
10 (cm) and the standing wave to be visible light, 
$\omega$ is larger than $c/\ell_1$ by about 6 orders. 
It means that the resonant increase can appear 
much earlier. 
Therefore, we can evaluate 
the applicable time of the linear approximation 
by using only the resonantly increasing term, {\it i.e.}, 
if $A_1, A_2\neq0$, 
\begin{equation}\label{appl_lin_appr}
t \ll \omega^{-1}\min 
\left(
\frac{A_1}{\sqrt{f_1^2+g_1^2}}, 
\frac{A_2}{\sqrt{p_1^2+q_1^2}}
\right). 
\end{equation}

\section{Leading part in long timescale}
We investigate the behavior of nonlinear electromagnetic wave 
in a longer timescale than the applicable range of linear 
approximation. 
We discuss a condition or timescale where 
the increase of a second harmonic wave can be excluded.  

A resonantly increasing part in Eq. (\ref{reso_mct_EB}) 
has the same spatial part as the corresponding 
classical term. Thus, if we calculate a higher-order 
correction by regarding the resonant part as a new wave source, 
the leading part of the higher-order correction should have 
the same spatial part. 
Therefore, we can expect that the leading part of the 
nonlinear electromagnetic wave has the same 
spatial part as the classical term, even in the longer timescale. 
We express such a leading part as 
\begin{equation}\label{def_EB_lp}
\begin{split}
&\boldsymbol{E}^{(\text{lp})}=
\begin{pmatrix}
-\sin \theta(f\sin T+g\cos T) \cos X \sin Y\\
\cos \theta (f\sin T+g\cos T) \sin X \cos Y\\
(p \sin T+q\cos T) \sin X \sin Y\\
\end{pmatrix}, \\
&\boldsymbol{B}^{(\text{lp})}=
\begin{pmatrix}
\sin \theta(-q \sin T+p \cos T) \sin X \cos Y\\
- \cos \theta(-q \sin T+p \cos T) \cos X \sin Y\\
(-g \sin T+f \cos T) \cos X \cos Y\\
\end{pmatrix}
+\boldsymbol{B}_s,\\
\end{split}
\end{equation}
where the envelope functions $f, g, p,$ and $q$ depend only on 
the normalized time $T$. 
Our concern is to verify the assumption of the leading part 
and to calculate these envelope functions. 
Note that the classical term is also 
included in the leading part. 
Thus, $\boldsymbol{E}^{(\text{lp})}$ and 
$\boldsymbol{B}^{(\text{lp})}$ are of the order of $A_1+A_2+B_s$. 
In particular, the envelope functions are 
of the order of $A_1+A_2$. 
All discarded terms are expected to be at most of the 
order of $(C_{2,0}+C_{0,2})(A_1+A_2+B_s)^3\ll A_1+A_2$. 
By substituting the leading part in the nonlinear 
Maxwell's equations in Eq. (\ref{eq_nme}), 
we can see that the envelope functions should 
satisfy the following differential equations: 
\begin{equation}\label{4_diff_eqs}
\begin{split}
&f'=\Gamma_1g+3\Gamma(f^2+g^2)g
+(\Gamma-\tilde{\Gamma})fpq
+\tilde{\Gamma}gp^2+\Gamma gq^2,\\
&g'=-\Gamma_1f-3\Gamma(f^2+g^2)f
-\Gamma fp^2-\tilde{\Gamma}fq^2
-(\Gamma-\tilde{\Gamma})gpq,\\
&p'=\Gamma_2q+3\Gamma(p^2+q^2)q
+(\Gamma-\tilde{\Gamma})fgp
+\tilde{\Gamma}f^2q+\Gamma g^2q,\\
&q'=-\Gamma_2p-3\Gamma(p^2+q^2)p
-(\Gamma-\tilde{\Gamma})fgq
-\Gamma f^2p-\tilde{\Gamma}g^2p.\\
\end{split}
\end{equation}
Note that these equations are closed with $f, g, p,$ and $q$. 
The initial values are 
\begin{equation}\label{ini_fgpq}
f(0)=A_1, g(0)=0, p(0)=A_2\cos \Phi, q(0)=A_2\sin \Phi. 
\end{equation}
We first check the appropriateness of the differential equations. 
By substituting the initial values in Eq. (\ref{ini_fgpq}) 
into the right-hand side of Eq. (\ref{4_diff_eqs}), 
the results agree with the right-hand side of Eq. (\ref{fgpq1}). 
Therefore, the solution of these differential equations 
can reproduce the resonant increase in the linear approximation. 
Furthermore, let 
\begin{equation}
\mathscr{X}=f^2+g^2+p^2+q^2, 
\end{equation}
we can see that 
the time differential of $\mathscr{X}$ 
is always zero and therefore, 
$\mathscr{X}=A_1^2+A_2^2$ is 
a conservative quantity. 
The conservative quantity indicates that 
the leading part of the total electromagnetic energy 
in the cavity is preserved. 
Because of these two properties, it will be reasonable 
to expect that the electromagnetic field in a long timescale 
is well approximated by Eq. (\ref{def_EB_lp}) and the 
time-evolution of $f, g, p,$ and $q$ is 
determined by Eq. (\ref{4_diff_eqs}). 
In the next section, we introduce a new function $\alpha$ 
to perform the analysis in an easier way. 
In section 8, we return to the equations by 
showing that $f, g, p,$ and $q$ are determined 
once $\alpha$ is obtained.

\section{Three differential equations for 
$\alpha, \beta,$ and $\gamma$}
\label{de_alpha_beta_gamma}

We introduce three functions $\alpha, \beta,$ and $\gamma$ as 
\begin{equation}
\begin{split}
&\alpha=\frac{1}{\mathscr{X}}(f^2+g^2),\\
&\beta=\frac{1}{\mathscr{X}}(fp+gq),\\
&\gamma=\frac{1}{\mathscr{X}}(fq-gp).\\
\end{split}
\end{equation}
We also introduce three constants as 
\begin{equation}
\begin{split}
&c_1=(\tilde{\Gamma}-\Gamma)\mathscr{X},\\
&c_2=(3\Gamma-\tilde{\Gamma})\mathscr{X},\\
&\xi=\Gamma_2-\Gamma_1+
(3\Gamma-\tilde{\Gamma})\mathscr{X}.\\
\end{split}
\end{equation}
Using Eq. (\ref{4_diff_eqs}), we obtain three 
differential equations for 
$\alpha, \beta,$ and $\gamma$ as 
\begin{equation}\label{3_diff_eqs_abc}
\begin{split}
&\alpha'=-2c_1\beta\gamma,\\
&\beta'=(\xi-2c_2\alpha)\gamma,\\
&\gamma'=-(\xi+c_1)\beta+2(c_1+c_2)\alpha\beta.\\
\end{split}
\end{equation}
Note that these differential equations are closed 
by the three functions. 
The initial values are given by 
\begin{equation}\label{ini_abc}
\alpha(0)=\frac{A_1^2}{\mathscr{X}}, \ \ 
\beta(0)=\frac{A_1A_2}{\mathscr{X}}\cos \Phi, \ \ 
\gamma(0)=\frac{A_1A_2}{\mathscr{X}}\sin \Phi. 
\end{equation}
It is clear from the definition that all 
$\alpha, \beta,$ and $\gamma$ are bounded and 
the right-hand sides of Eq. (\ref{3_diff_eqs_abc}) are proven to be 
Lipschitz continuous. Therefore, 
a unique solution exists for the initial values. 

We introduce a constant of integration $Z$ 
by integrating the second line in Eq. (\ref{3_diff_eqs_abc}) 
after multiplying by $-2c_1\beta$. 
Let 
\begin{equation}
\begin{split}
&P_1(\alpha)=c_2\alpha^2-\xi\alpha-Z,\\
&P_2(\alpha)=-(c_1+c_2)\alpha^2
+(\xi+c_1)\alpha+Z,\\
\end{split}
\end{equation}
we obtain 
\begin{equation}\label{beta_gamma_P1P2}
c_1\beta^2=P_1(\alpha), \ \ c_1\gamma^2=P_2(\alpha). 
\end{equation}
We can also see 
\begin{equation}\label{P1+P2}
P_1(\alpha)+P_2(\alpha)=c_1\alpha(1-\alpha), 
\end{equation}
and 
\begin{equation}\label{a2P_1P_2}
\alpha'^2=4P_1(\alpha)P_2(\alpha). 
\end{equation}
We can immediately obtain $\alpha$ from the last equation 
if the sign of $\alpha'$ does not change. 
Even though the sign of $\alpha'$ can change, 
we can obtain a second order differential equation 
that only includes $\alpha$, which must be easier to solve than 
Eq. (\ref{3_diff_eqs_abc}). 
Therefore, we focus on $\alpha$. 
As we show later, the leading part of the electromagnetic 
field can be obtained once $\alpha$ is calculated.

Because of Eq. (\ref{a2P_1P_2}), 
the range of possible $\alpha$ is limited to satisfy 
$P_1P_2\ge 0$. 
Thus, we assign symbols for the roots of 
$P_1$ and $P_2$. 
They are given by 
\begin{equation}\label{}
\begin{split}
&\alpha_{1\pm}=\frac{\xi}{2c_2}\pm \frac{1}{2c_2}
\sqrt{\xi^2+4c_2Z},\\
&\alpha_{2\pm}=\frac{\xi+c_1}{2(c_1+c_2)}\pm 
\frac{1}{2(c_1+c_2)}
\sqrt{(\xi+c_1)^2+4(c_1+c_2)Z},\\
\end{split}
\end{equation}
respectively.

\section{Range of parameters $c_1,c_2,\xi,$ and $Z$}

Since the parameters $c_1$ and $c_2$ appear in the 
coefficients of 
the highest degrees of $P_1$ and $P_2$, 
their signs are especially important. 
In the Heisenberg-Euler model, 
they are positive as 
$c_1 \ge (1/8)C_{2,0}\mathscr{X}>0$ and 
$c_2=(5/8)C_{2,0}\mathscr{X}>0$. 
Thus, we fix them to be positive constants, {\it i.e.}, 
we fix the nonlinear vacuum model,  
$\mathscr{X}=A_1^2+A_2^2,$ and $\theta$. 
In this case, $P_1$ 
is convex downward and 
$P_2$ is convex upward. 
Equation (\ref{P1+P2}) shows that 
$\alpha \neq 0,1$ is never a common root of 
$P_1$ and $P_2$.

\begin{figure}
\begin{center}
\begin{tabular}{c}
\includegraphics[width=8cm]{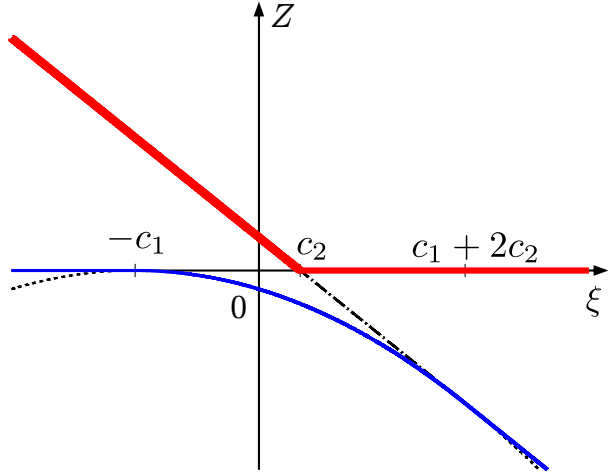} \\
\end{tabular}
\end{center}
\caption{The range of possible $Z$ for each $\xi$ 
indicated in Eq. (\ref{Z_range_for_xi}). 
The upper and lower limits are shown by the bold red 
and thin blue curves, respectively. 
The dotted curve and chained line express 
$Z=-(\xi+c_1)^2/[4(c_1+c_2)]$ and 
$Z=-\xi+c_2$, respectively. 
\label{fig_Z_xi}}
\end{figure}

We regard $\xi$ to be a variable parameter 
since it is controllable by the 
value of the static magnetic flux density. 
Further, $Z$ is treated as another variable parameter 
because the initial values depend only on it. 
Therefore, it is farsighted to classify the 
time-evolution by the values of $\xi$ and $Z$. 
The possible range of $Z$ for each $\xi$ is given by 
\begin{equation}\label{Z_range_for_xi}
\begin{cases}
0\le Z \le -\xi+c_2 & (\xi \le -c_1)\\
-\frac{(\xi+c_1)^2}{4(c_1+c_2)}
\le Z \le -\xi+c_2
& (-c_1 \le \xi \le c_2)\\
-\frac{(\xi+c_1)^2}{4(c_1+c_2)}
\le Z \le 0 
& (c_2 \le \xi \le c_1+2c_2)\\
-\xi+c_2 \le Z \le 0 
& (c_1+2c_2 \le \xi).\\
\end{cases}
\end{equation}
The region is shown in Fig. \ref{fig_Z_xi}.

\section{Classification of $\alpha$ for every $\xi$ and $Z$}

There are three possible behaviors of $\alpha$, {\it i.e.}, 
keeps oscillating, converges to a certain value, and 
remains as the initial value. 
The values of $\xi, Z,$ and $\alpha(0)$ determine the behavior.

\begin{figure}
\begin{center}
\begin{tabular}{c}
\includegraphics[width=12cm]{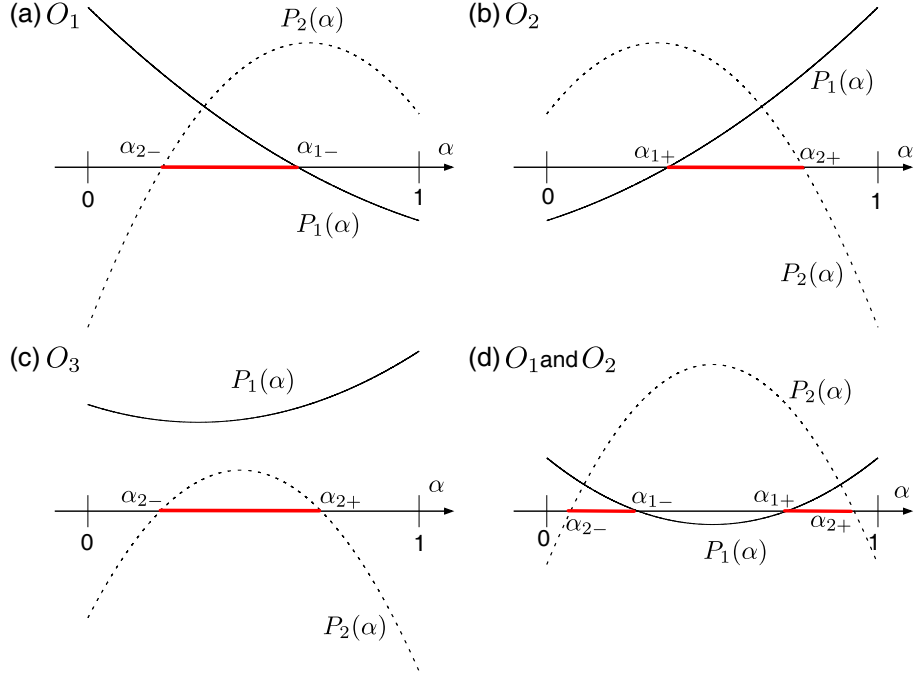} \\
\end{tabular}
\end{center}
\caption{
Typical behaviors of 
$P_1$ and $P_2$ 
for the possible 
subtypes of oscillating $\alpha$. 
The zeros of $P_1$ or $P_2$ 
correspond to 
the maximum and minimum values of $\alpha$ 
for each subtype. 
The range of $\alpha$ is 
highlighted by the bold red lines. 
Both subtypes $O_1$ and $O_2$ are possible 
in the case of (d). 
\label{P1P2_123_12}}
\end{figure}

In the case of oscillation, since 
$P_1P_2\ge 0$ is necessary, 
we can classify the maximum value $\alpha_M$ 
and minimum value $\alpha_m$ 
in the following three subtypes $O_1, O_2,$ and $O_3$ 
given in Table \ref{table_aMam}. 
Typical $P_1$ and $P_2$ 
for each subtype 
are shown in Fig. \ref{P1P2_123_12}.

\begin{table}
\centering
\begin{tabular}{ccc}
\hline
Subtype & $\alpha_m$ & $\alpha_M$ \\
\hline \hline
$O_1$ & $\alpha_{2-}$ & $\alpha_{1-}$ \\
$O_2$ & $\alpha_{1+}$ & $\alpha_{2+}$ \\
$O_3$ & $\alpha_{2-}$ & $\alpha_{2+}$ \\
\hline
\end{tabular}
\caption{The minimum and maximum values 
for each subtype of oscillating $\alpha$. 
\label{table_aMam}}
\end{table}

The subtype $O_1$ 
is realized if 
\begin{equation}\label{shindou_A}
\begin{cases}
-\xi+c_2<Z<0 
& (c_2<\xi) \\
Z=-\xi+c_2
& (c_2<\xi<2c_2)\\
-\frac{\xi^2}{4c_2}<Z<
\min(0,-\xi+c_2)
& (0<\xi<2c_2), \\
\end{cases}
\end{equation}
and $\alpha_{2-}\le \alpha(0) \le \alpha_{1-}$ holds. 
In the first case, the condition for the initial value 
is automatically satisfied. 
The subtype $O_2$ 
is realized if 
\begin{equation}\label{shindou_B}
\begin{cases}
0<Z<-\xi+c_2
& (\xi<c_2) \\
Z=0 & (0<\xi<c_2) \\
-\frac{\xi^2}{4c_2}<Z<\min(0,-\xi+c_2)
& (0<\xi<2c_2), \\
\end{cases}
\end{equation}
and $\alpha_{1+}\le \alpha(0) \le \alpha_{2+}$ holds. 
In the first case, the condition for the initial value 
is automatically satisfied. 
The subtype $O_3$ 
is realized if 
\begin{equation}\label{shindou_C}
\begin{cases}
-\frac{(\xi+c_1)^2}{4(c_1+c_2)}<Z<0 
& (-c_1 < \xi \le 0) \\
-\frac{(\xi+c_1)^2}{4(c_1+c_2)}<Z<-\frac{\xi^2}{4c_2}
& (0<\xi<2c_2) \\
-\frac{(\xi+c_1)^2}{4(c_1+c_2)}<Z<-\xi+c_2
& (2c_2 \le \xi < c_1+2c_2), \\
\end{cases} 
\end{equation}
where 
$\alpha_{2-} \le \alpha(0) \le \alpha_{2+}$ 
holds necessarily.

In the case that $\alpha$ converges, 
the limit value, initial condition, and range of 
$\xi$ and $Z$ are given by 
\begin{equation}\label{jouken_syuusoku}
\begin{cases}
\alpha\to0 & \left( 
\alpha(0)\neq 0 \land 
-c_1<\xi \le 0 \land Z=0 
\right)\\
\alpha\to 1 & \left( 
\alpha(0)\neq 1 \land 
2c_2\le \xi<c_1+2c_2 \land 
Z=-\xi+c_2 
\right)\\
\alpha\to \frac{\xi}{2c_2} & \left( 
\alpha(0)\neq \frac{\xi}{2c_2} \land 
0<\xi<2c_2 \land Z=-\frac{\xi^2}{4c_2} 
\right),\\
\end{cases}
\end{equation}
where the symbol $\land$ is a logical conjunction. 

$\alpha$ does not change 
if its initial value and the range of $\xi$ and $Z$ satisfy 
\begin{equation}\label{alpha_ugokanai}
\begin{cases}
\alpha(0)=1 & \left(\forall \xi \land Z=-\xi+c_2 
\right)\\
\alpha(0)=0 & \left(\forall \xi \land Z=0 
\right)\\
\alpha(0)=\frac{\xi}{2c_2}  & \left(
0 < \xi < 2c_2 \land Z=-\frac{\xi^2}{4c_2} \right)\\
\alpha(0)=\frac{\xi+c_1}{2(c_1+c_2)}  & \left(
-c_1 < \xi < c_1+2c_2 \land 
Z=-\frac{(\xi+c_1)^2}{4(c_1+c_2)} \right).\\
\end{cases}
\end{equation}

We have completely classified the behavior of $\alpha$ 
for all possible $\xi$ and $Z$. 
If $\alpha$ is constant, there is nothing to do. 
If $\alpha$ converges, $\alpha'$ can change its sign 
at most once. Thus, all we have to do is to 
integrate the adequate square root of Eq. (\ref{a2P_1P_2}) 
at most twice. 
In the case of oscillating $\alpha$, 
the generalization of Jacobi's elliptic function is given 
in Appendix A and the solution can be obtained by 
a change of variable.

Note that if 
\begin{equation}\label{AB_Z}
0 < \xi < 2c_2 \land 
-\frac{\xi^2}{4c_2}<Z<\min(0,-\xi+c_2), 
\end{equation}
then both subtypes 
$O_1$ 
and 
$O_2 $ 
are possible because the roots of $P_1$ and $P_2$ satisfy 
\begin{equation}
0<\alpha_{2-}<\alpha_{1-}<\alpha_{1+}<\alpha_{2+}<1. 
\end{equation}
This is shown in Fig. \ref{P1P2_123_12}(d). 
The initial value $\alpha(0)$ determines 
which subtype is realized.

\section{Sufficiency of obtaining $\alpha$}

Here we have solved $\alpha$. If it changes with time, 
it always satisfies $0<\alpha<1$. 
Let 
\begin{equation}
\begin{split}
&\varphi_1(T)=-(\Gamma_2+3\Gamma \mathscr{X})T
-Z\int_0^T\frac{\text{d}\tau}{\alpha(\tau)},\\
&\varphi_2(T)=\Phi-(\Gamma_1+3\Gamma \mathscr{X})T
-(\xi-c_2+Z)
\int_0^T\frac{\text{d}\tau}{1-\alpha(\tau)},\\
\end{split}
\end{equation}
the four envelope functions are given as 
$f=\sqrt{\alpha \mathscr{X}}\cos \varphi_1, 
g=\sqrt{\alpha \mathscr{X}} \sin \varphi_1,
p=\sqrt{(1-\alpha)\mathscr{X}} \cos \varphi_2,$ and 
$q=\sqrt{(1-\alpha)\mathscr{X}} \sin \varphi_2$. 
Substituting them into Eq. (\ref{def_EB_lp}) leads to 
\begin{equation}
\begin{split}
\boldsymbol{E}^{(\text{lp})}=&
\sqrt{\alpha \mathscr{X}}
\begin{pmatrix}
- \sin \theta \cos X \sin Y\\
 \cos \theta \sin X \cos Y\\
0\\
\end{pmatrix}
\sin(T+\varphi_1)\\
&+\sqrt{(1-\alpha)\mathscr{X}}
 \sin X \sin Y\sin(T+\varphi_2)\boldsymbol{e}_z,\\
\end{split}
\end{equation}
where $\boldsymbol{e}_z$ is the unit vector of 
the $z$ direction. 
We can interpret $\varphi_1$ and $\varphi_2$ as 
the phase changes of each mode 1 and 2, respectively. 
The relative phase can be defined by $\varphi_2-\varphi_1$. 
On the contrary to the nonlinear standing wave 
in a one-dimensional cavity 
\cite{PhysRevA.104.063513}, 
the relative phase continuously changes with time.

Since the analysis has been completed essentially, we describe 
the result in the physical context. 
We can interpret $\alpha$ to express the intensity ratio between 
the two modes. 
Our analysis reveals that there are three possible types 
for the time evolution of $\alpha$, 
depending on the nonlinear parameters, 
direction of the standing waves, and 
static magnetic flux density. 
In the first type, the energy transfer continuously 
occurs between the two modes. The maximum and minimum 
ratios are further classified into three subtypes. 
In the second type, the energy transfer eventually 
decreases to zero and the mode ratio converges. 
In the third type, each mode keeps its initial energy.

\section{Examples of $\alpha$}
We give example of $\alpha$ for the oscillating and 
converging types, as well as the electric field for a constant 
$\alpha$.

\begin{figure}
\begin{center}
\begin{tabular}{c}
\includegraphics[width=14cm]{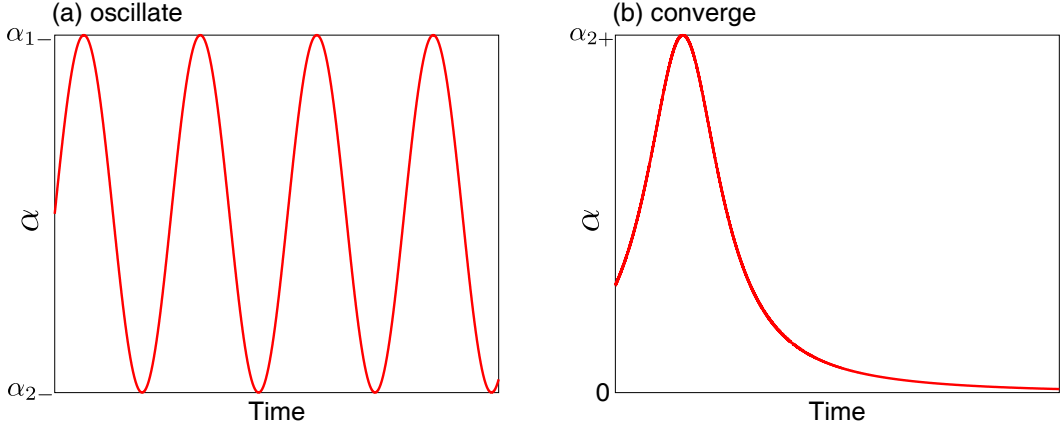} \\
\end{tabular}
\end{center}
\caption{Examples for oscillating and converging $\alpha$. 
(a) The oscillating subtype $O_1$ is demonstrated. 
$\xi$ and $Z$ satisfy the third case in 
Eq. (\ref{shindou_A}). 
The initial value is set to 
$\alpha(0)=(\alpha_{1-}+\alpha_{2-})/2$ 
to meet the condition. 
(b) The converging type is demonstrated with 
the first case of Eq. (\ref{jouken_syuusoku}) and 
$\alpha'(0)>0$. 
\label{alpha_shindou_I_shuusoku}}
\end{figure}

\subsection{Oscillating $\alpha$}
In the case that $\alpha$ oscillates, 
it is given by 
\begin{equation}\label{kai_sindou_alpha}
\alpha=\frac{\alpha_M+\alpha_m}{2}
+\frac{\alpha_M-\alpha_m}{2}\mathcal{S}(\sqrt{\mathscr{A}}T+T_0;\lambda_1,\lambda_2), 
\end{equation}
where the function $\mathcal{S}$ is a generalization of 
Jacobi's elliptic function $\text{sn}$ described in 
Appendix A. 
The constants 
$\mathscr{A}, \lambda_1,$ and $\lambda_2$ are 
determined by the oscillation subtype. 
$T_0$ depends on the oscillation subtype and the initial value. 
They are given in Appendix B.

\subsection{Converging $\alpha$}

As aforementioned, we can obtain $\alpha$ 
simply by integrating an adequate square root of 
Eq. (\ref{a2P_1P_2}). For the first case in 
Eq. (\ref{jouken_syuusoku}), 
$\alpha$ converges to zero. 
Let $\bar{a}>0$ be a certain constant, 
$\alpha$ behaves at sufficiently large $T$ as
\begin{equation}\label{}
\alpha \approx \bar{a}
\exp\left[-4(c_1+c_2)c_2
\sqrt{-\alpha_{1-}\alpha_{2+}}T\right], 
\end{equation}
converges to zero.

\subsection{Constant $\alpha$}
If $\alpha$ is constant, the amplitudes of 
two modes are also constant. 
As examples, we show the solutions for the 
first and third cases of Eq. (\ref{alpha_ugokanai}). 

$A_2=0$ corresponds to the first case. 
Let 
\begin{equation}\label{Omega1}
\Omega_1=\Gamma_1+3\Gamma A_1^2, 
\end{equation}
we obtain $\varphi_1(T)=-\Omega_1T$ and 
\begin{equation}
\boldsymbol{E}^{(\text{lp})}=
A_1
\begin{pmatrix}
-\sin \theta \cos X \sin Y\\
\cos \theta \sin X \cos Y\\
0\\
\end{pmatrix}
\sin(T-\Omega_1T). 
\end{equation}

The third case is realized if and only if 
$A_1>0, A_2>0, \cos\Phi=0,$ and 
\begin{equation}
(12C_{2,0}-C_{0,2})(A_1^2-A_2^2)
=8(4C_{2,0}-C_{0,2})
(B_{sx}^2\sin^2\theta+B_{sy}^2\cos^2\theta-B_{sz}^2). 
\end{equation}
Let 
\begin{equation}\label{Omega3}
\Omega_3=\Gamma_2+3\Gamma \mathscr{X}-\frac{\xi}{2}, 
\end{equation}
we obtain 
$\varphi_1(T)=-\Omega_3T,
\varphi_2(T)=\Phi-\Omega_3T,$ and 
\begin{equation}
\begin{split}
\boldsymbol{E}^{(\text{lp})}=&
A_1
\begin{pmatrix}
-\sin \theta \cos X \sin Y\\
\cos \theta \sin X \cos Y\\
0\\
\end{pmatrix}
\sin(T-\Omega_3 T)\\
&+A_2\sin \Phi
\sin X \sin Y \cos(T-\Omega_3T)\boldsymbol{e}_z.\\
\end{split}
\end{equation}

\section{Discussion}

We have demonstrated in the linear approximation 
that the uniform current generates an increasing 
correction of the second harmonic wave 
if $\cos k\ell_1=0$ or $\cos k\ell_2=0$ holds. 
It is of interest to know whether the second harmonic 
is still generated when the leading part of the electromagnetic 
field differs from the classical term. To answer the question, 
it is sufficient to calculate a uniform current 
by applying the linear approximation for the leading part. 
The result is given by 
\begin{equation}\label{uni_j_lp}
\tilde{\boldsymbol{j}}_{\text{uni}}=-\frac{k}{4}
(4C_{2,0}-C_{0,2})\mathscr{X}
\sin 2\theta 
\sqrt{\alpha(1-\alpha)}
\cos(2T+\varphi_1+\varphi_2)
\begin{pmatrix}
B_{sy}\\
B_{sx}\\
0\\
\end{pmatrix}.
\end{equation}
Equation (\ref{j_Ckl}) is reproduced 
if the initial values are substituted. 
The time variations of 
$\alpha, \varphi_1,$ and $\varphi_2$ 
are sufficiently slower than the period of one cycle 
$2\pi/\omega$ and 
thus we can regard them to be constant in such a short timescale. 
Therefore, if $\alpha$ does not rapidly converge to zero or unity, 
the second harmonic will keep increasing 
and become comparable to the classical term 
in the course of time. 
This consideration suggests that 
Eq. (\ref{def_EB_lp}) is a good approximation 
in the timescale of 
$t\lesssim \ell_1(A_1+A_2)/
[c(C_{2,0}+C_{0,2})(A_1+A_2+B_s)^3]$. 
This value is, of course, much larger than 
the right-hand side in Eq. (\ref{appl_lin_appr}). 
Therefore, the validity of leading part calculation is unshaken.

\section{Final remarks}

We have analyzed the nonlinear electromagnetic wave 
in the two-dimensional rectangular cavity. 
The classical electromagnetic field is given as 
the two modes of standing wave and 
a constant magnetic flux density. 
Applying the linear approximation to calculate the 
corrective term in a short timescale, 
the result suggests that the leading part of the nonlinear 
electromagnetic wave has the same spatial distribution 
as the classical standing wave. 
Using this supposition, we derived nonlinear 
simultaneous differential equations 
which express the time evolution of the leading part. 
We introduced a new variable $\alpha$ which expresses 
the intensity ratio of the two modes. 
The behavior of $\alpha$ is classified into three types,
{\it i.e.}, keep oscillating, converging, and constant. 
Once $\alpha$ is obtained, 
the leading part of the nonlinear electromagnetic wave 
can be calculated immediately. 
It should be noted that the behaviors of the leading part in a 
long timescale is completely different between the 
two- and one-dimensional cavities. If we take 
the limits of $\cos Y\to1$ and $\cos \theta\to1$, 
the leading part in Eq. (\ref{def_EB_lp}) 
does not converge to the corresponding 
one-dimensional solution \cite{PhysRevA.104.063513} 
while the classical term in 
Eq. (\ref{koten_EcBc}) converges to 
a one-dimensional standing wave.

In the applicable range of linear approximation, 
we found a characteristic feature which has not appeared in 
the one-dimensional cavity system. 
In the two-dimensional cavity, the uniform current 
given in Eq. (\ref{j_Ckl}) or 
Eq. (\ref{uni_j_lp}) can yield an increase 
of corresponding corrective term, 
if the cavity size satisfies $\cos k\ell_1=0$ or $\cos k\ell_2=0$.  
The increased electromagnetic wave is a 
one-dimensional second harmonic. 
It increases with time but is not proportional to time, 
resulting in a slower growth than the resonant increase.

\begin{acknowledgments}
The author thanks 
Dr. M. Nakai and Dr. K. Mima 
for discussions on the nonlinear 
electromagnetic wave and its experimental application.
The author quite appreciates Dr. J. Gabayno for checking 
the logical consistency of the text. 
\end{acknowledgments}


%

\end{document}